\documentclass[prd,showpacs,preprintnumbers,nofootinbib,amsmath,eqsecnum,twocolumn]{revtex4}
 \usepackage[dvips,final]{graphicx}
  \usepackage{amssymb}
   \usepackage{amsmath}
    \usepackage{epsfig}
     \usepackage{bm}% bold math
      \usepackage{pifont}
\def\Li{\relax\ifmmode{\textbf{Li}_{2}}\else{Li$_2${ }}\fi}

\newcommand{\be}{\begin{equation}}
\newcommand{\ee}{\end{equation}}

\newcommand{\ba}{\begin{eqnarray}}
\newcommand{\ea}{\end{eqnarray}}
\newcommand{\bg}{\begin{gather}}
\newcommand{\foma}{\end{gather}}

\newcommand{\vecc}[1]{\mbox{\boldmath $#1$}}

\def\<{\langle}
\def\>{\rangle}

\def\({\left(}
\def\[{\left[}
\def\){\right)}
\def\]{\right]}

\begin{document}
\thispagestyle{empty}

\title{Evolution of cusped light-like Wilson loops and geometry of the loop space}
\author{I.O.~Cherednikov}
\email{igor.cherednikov@ua.ac.be}
\affiliation{Departement Fysica, Universiteit Antwerpen, B-2020 Antwerpen, Belgium\\
and\\
BLTP JINR, RU-141980 Dubna, Russia}
\author{T.~Mertens}
\email{tom.mertens@ua.ac.be}
\affiliation{Departement Fysica, Universiteit Antwerpen, B-2020 Antwerpen, Belgium\\}
\author{F.F.~Van der Veken}
\email{frederik.vanderveken@ua.ac.be}
\affiliation{Departement Fysica, Universiteit Antwerpen, B-2020 Antwerpen, Belgium\\}
\vspace {10mm}
%\cleardoublepage
\date{\today} 

\begin{abstract}
We discuss the possible relation between certain geometrical properties of the loop space and energy evolution of the cusped Wilson exponentials defined on the light-cone. Analysis of the area differential equations for this special class of the Wilson loops calls for careful treatment of the ultraviolet and rapidity divergences which make those loops non-multiplicatively-renormalizable. We propose to consider the renormalization properties of the light-cone cusped Wilson loops from the point of view of the universal quantum dynamical approach introduced by Schwinger. We discuss the relevance of the Makeenko-Migdal loop equations supplied with the modified Schwinger principle to the energy evolution of some phenomenologically significant objects, such as transverse-momentum dependent distribution functions, collinear parton densities at large-$x$, etc.
\end{abstract}
%Keywords: Gauge invariance
%          Wilson lines and loops
%          Renormalization and anomalous dimensions
\maketitle

%\tableofcontents

%%%%%%%%%%%%%%%%%%%%%%%%%%%%%%%%%%%%%%%%%%%%%%%%%%%%%%%%%%%%%%%%%%%%%%%
%%%%%%%%%%%%%%%%%%%%%%%%%%%%%%%%%%%%%%%%%%%%%%%%%%%%%%%%%%%%%%%%%%%%%%%
%\cleardoublepage

\section{Introduction}
\label{sec:intro}
Complete recast of QCD in the loop space would enable us to use colorless gauge-invariant field functionals as the fundamental degrees of freedom instead of the colored gauge-dependent quarks and gluons \cite{Loop_Space, WL_RG}. The physical observables are supposed to be expressed via the vacuum averages of the Wilson loops depending, in general, on multiple contours $\{\Gamma_i\}$:
\begin{eqnarray}
 & & W_n (\Gamma_1, ... \Gamma_n)
  = \nonumber \\
& & \Big \langle 0 \Big| {\cal T} \frac{1}{N_c} {\rm Tr}\ \Phi (\Gamma_1)\cdot \cdot \cdot \frac{1}{N_c}{\rm Tr}\ \Phi (\Gamma_n)  \Big| 0 \Big\rangle \ , \label{eq:wl_def} \\
 & & \Phi (\Gamma_i)
   =
   {\cal P} \ \exp\[ig \oint_{\Gamma_i} \ dz^\mu A_{\mu} (z) \] \ . \nonumber
\end{eqnarray}
Here and in what follows the gauge fields $A_\mu$ belong to the fundamental representation of non-Abelian gauge group  $SU(N_c)$.
Generically, dynamics of any reasonable function of the gauge fields is properly determined by the set of the Schwinger-Dyson equations in the following form:
\begin{equation}
 \langle 0 | \nabla_\mu F^{\mu\nu} \  | 0 \rangle
=
i \langle 0 |  \frac{\delta}{\delta A_\nu} \  | 0 \rangle \ .
\label{eq:sch_dy_YM}
\end{equation}
Being applied to the scalar functionals $\Phi (\Gamma)$ (\ref{eq:wl_def}) which constitute the loop space, Eq. (\ref{eq:sch_dy_YM}) results in the Makeenko-Migdal (MM) equations \cite{MM_WL}:
\begin{eqnarray}
 & & \partial_x^\nu \ \frac{\delta}{\delta \sigma_{\mu\nu} (x)} \ W_1(\Gamma)
 = \nonumber \\
 & & N_c g^2 \ \oint_{\Gamma} \ dz^\mu \ \delta^{(4)} (x - z) W_2(\Gamma_{xz} \Gamma_{zx}) \ ,
 \label{eq:MM_general}
\end{eqnarray}
where the key operations are the area $\delta/\delta\sigma_{\mu\nu} (x)$ and the path $\partial_\mu (x)$ derivatives \cite{MM_WL}:
\begin{equation}
 \frac{\delta}{\delta \sigma_{\mu\nu} (x)} \ \Phi (\Gamma)
 \equiv
 \lim_{|\delta \sigma_{\mu\nu} (x)| \to 0} \ \frac{ \Phi (\Gamma\delta \Gamma) - \Phi (\Gamma) } {|\delta \sigma_{\mu\nu} (x)|} \ ,
\label{eq:area_derivative}
\end{equation}
where the contour $\Gamma\delta \Gamma$ is obtained from the initial one, $\Gamma$, after an infinitesimal area deformation $\delta \Gamma$ at some point $x$, while the infinitesimal {\it pinch} of the path $\Gamma$ at the point $x$, without changing its area, makes it possible to introduce the single-point path derivative
\begin{equation}
 \partial_\mu  \Phi(\Gamma)
 =
 \lim_{|\delta x_{\mu}|} \frac{\Phi(\delta x_\mu^{-1}\Gamma\delta x_\mu) - \Phi(\Gamma)}{|\delta x_{\mu}|} \ .
 \label{eq:path_derivative}
\end{equation}
Note that, alternatively, the area derivative can be written in the Polyakov form
\begin{equation}
 \frac{\delta}{\delta \sigma_{\mu\nu} [x(\tau)]}
 =
 \int_{\tau+0}^{\tau-0}\! d\tau' (\tau' - \tau)
 \ \frac{\delta}{\delta x_\mu (\tau')} \frac{\delta}{\delta x_\mu (\tau)} \ .
\end{equation}
The latter definition had been adopted in, e.g., \cite{St_Kr_WL_cast} to approach the similar problems from a different point of view.

The standard way of derivation of the loop equations within the general Schwinger-Dyson framework is based on the Mandelstam formula
\begin{equation}
 \frac{\delta}{\delta \sigma_{\mu\nu} (x)} \ \Phi (\Gamma)
 =
 ig {\rm Tr} \[ F_{\mu\nu} \ \Phi (\Gamma_x)  \]
\end{equation}
and/or utilizes the Stokes theorem. This approach is certainly relevant to the class of smooth Wilson loops without cusps\footnote{We appreciate clarifying discussions with I.V. Anikin on this and related topics.}.
The MM equations in the form (\ref{eq:MM_general}) are exact and nonperturbative and reflect the differential geometrical structure of the loop space. However, the range of practical use of the MM equations is quite restricted because of the following reasons \cite{WL_Renorm}. First, most physically interesting Wilson loops develop ultraviolet, infrared and light-cone divergences; in addition to that, they possess specific obstructions, cusps and/or self-intersections, which yield yet other problems with corresponding divergences. It is known, however, that Eq. (\ref{eq:MM_general}) can not be applied straightforwardly to the lightlike Wilson loops with cusps. The renormalized version of the MM equation which is valid for these loops is not available. One of the reasons is that the area functional derivative is not a well-defined operation for arbitrary contours with obstructions.

Next, there are subtle points related to the continuous deformations of the paths in Minkowski space-time making the meaning of the derivatives unclear. In particular, in \cite{Topo_ST} it is argued that Minkowski space-time is as unconnected as a space can be with respect to  a path-topology (this resembles the structure of the set of rational numbers in the space of real numbers: although the rational numbers are infinitely close to each other, one cannot move from a rational number to another one without crossing a real number). Several attempts to define a correct path/loop space in order to solve this problem have been made without success. Recently, the new developments in the field of twistor theory have shown the MM equations to be valid, but the calculations are implemented in a completely different background (twistor space) \cite{WL_TT_ScA_2011}. General solutions of the MM equations in the four-dimensional space-time are also not known.

On the other hand, there are several approximations and simplifications which might make life easier. First,
in the 't Hooft large-$N_c$ limit, the factorization property allows one to get the MM Eq. closed: $W_2 (C_1,C_2) \approx W_1(C_1) \cdot W_1 (C_2)$ \cite{MM_WL}. Second, restricting ourselves to only light-cone rectangular contours, we end up with an effectively two-dimensional case: the space-time dimension where there is hope to solve the MM Eqs. Further, if one concentrates on the light-like polygons, then the angles between light-rays are fixed and conserved under any allowed area or path variation (that imposes additional constraint on the possible variations, of course). Hence, there are no angle-dependent contributions which may make the area differentiation ill-defined. Finally, the power of divergency decreases under the area differentiation, making it possible to construct appropriate renormalization-group equations for those loops.

In this work, we address some of these issues, considering a special class of the Wilson loops, namely the rectangular quadrilaterals with their sides being defined on the light cone. Strong interest to the cusped lightlike polygonal eikonal paths is related to the recently observed duality between the $n-$gluon planar scattering amplitudes in the ${\cal N} = 4 $ super-Yang-Mills theory and the vacuum average of the Wilson loops formed, correspondingly, by $n$ light-like segments connecting the space-time points ${x_i}$, so that their {\it lengths} $x_i - x_{i+1} = p_i$ are set equal to the external momenta of the $n-$gluon amplitude (see, e.g., \cite{WL_CFT} and references therein). It has been shown that the infrared evolution of the former is (or is expected to be) dual to the ultraviolet evolution of the latter, with the the cusp anomalous dimension being the main ingredient of the evolution equations \cite{KR87}. Therefore, the dynamical content, for instance, of a $2 \to 2$ scattering process in the momentum space maps the local geometrical properties of the light-like quadrilateral Wilson loop defined in the coordinate space. The local properties of Minkowskian paths in vicinity of the obstructions are known to be expressed in terms of the universal path-independent cusp anomalous dimension.

Further, the Wilson loops containing light-like segments were studied a couple of decades ago within a different context \cite{WL_LC_rect}. It was demonstrated that the renormalization properties of these objects are more complicated than the renormalization-group behavior of the cusped Wilson loops off the light-cone. In particular, the light-cone Wilson loops are not multiplicatively renormalizable due to the peculiar light-cone singularities arising in addition to the common ultraviolet and the infrared ones. Still it is possible to construct a combined renormalization-group equation reckoning with both the ultraviolet and the light-cone divergences, so that its solution does not show any pathological behavior. The cusp anomalous dimension which enters this equation is known to be of remarkable universality: it controls, for example, the infrared asymptotic behavior of phenomenologically important quantities such as the QCD and QED Sudakov form factors, the gluon Regge trajectory, the integrated (collinear) parton distribution functions at large-$x$, the anomalous dimension of the heavy quark effective theory, etc. \cite{WL_LC_rect, KR87, CAD_universal, St_Kr_Pheno}.

On the other hand, recent study of the operator definition of transverse-momentum dependent parton densities (TMDs) reveals that these  quantities, taken literally, demonstrate the similar extra light-cone singularities associated with rapidity divergences \cite{WL_LC_rapidity, CS_all}. The virtual Feynman graphs producing such terms are shown in Fig.\ 1.

\begin{figure}[ht]
 $$\includegraphics[angle=90,scale=0.5]{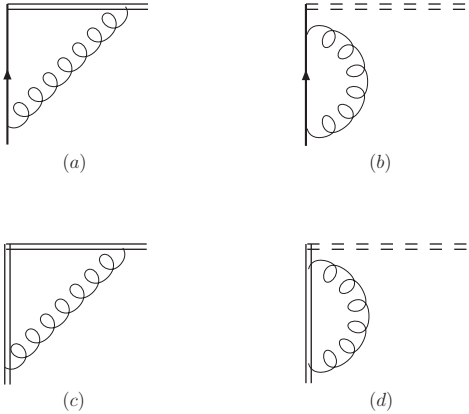}$$
   \vspace{0.0cm}
   \caption{\label{fig:1}The virtual one-loop Feynman graphs which produce extra singularities: $(a)$---vertex-type fermion-Wilson line interaction in covariant gauge; $(b)$---self-energy graph which yields the extra divergences in the light-cone gauge; $(c,d)$ are the counter-parts of $(a,b)$ from the soft factor made of the Wilson lines.}
\end{figure}
Therefore, it is instructive to undertake a detailed study of those properties shared by these apparently different quantities which originate in their light-cone structure and reveal themselves in the {\it too much singular} terms as compared to the off-light-cone objects.

\section{Dynamical loop equations, area development and energy evolution}
Evaluation of the cusped light-cone Wilson loops beyond the tree approximation in different (covariant, axial, light-cone, contour, etc.) gauges and corroboration of gauge independence of the result demands careful treatment of different classes of divergences even in the one-loop order. Special attention must by paid to the separation of the light-cone (also known as rapidity) singularities and the normal ultraviolet poles. We refer for more detailed discussion and further references to \cite{WL_RG, WL_LC_rect, WL_LC_rapidity, LC_integrals}.
In the large-$N_c$ limit we have in the coordinate space \cite{WL_LC_rect}
\begin{eqnarray}
& & W(\Gamma_\Box)  = 1 - \frac{1}{\epsilon^2}\ \frac{\alpha_s N_c}{2\pi}  \times  \nonumber \\
& & \(\[{-2 N^+ N^-\mu^2 + i0}\]^\epsilon + \[{2 N^+ N^-\mu^2 + i0}\]^\epsilon  \) \\ & &  + \frac{\alpha_s N_c}{2\pi} \( \frac{1}{2} \ln^2 \frac{N^+N^-}{-N^+N^-} + {\rm finite\ terms}  \) + O(\alpha_s^2) \ , \nonumber
\label{eq:WL_LC_1loop}
\end{eqnarray}
where the energy variables in the momentum space ${ s = (p_1 + p_2)^2}, {t = (p_2 + p_3)^2}$ match the geometrical {area} variables in the coordinate null-plane ${ s/2} = {- t/2} \to N^+ N^-$.
We will show in a separate paper that the result (\ref{eq:WL_LC_1loop}) is not only gauge invariant, but remarkably independent of any regularization of light-cone and ultraviolet divergences and of the way of their separation \cite{ChMVdV_2012}. This issue is of particular importance to the understanding of the operator structure of transverse-momentum dependent parton densities and soft-collinear effective theory (see, e.g., \cite{CS_all, SCET_TMD} and references therein). What happens to the issue of regularization-independence in the next-to-leading order is not known yet and deserves its own dedicated study.

Let us fix the transverse null-plane by imposing the condition $\vecc z_\perp = 0$; therefore, the area differentials are well-defined:
\begin{eqnarray}
& & { \delta \sigma^{+-} }
=
{ N^+ \delta N^- } \to p_1 \delta p_2 = \frac{1}{2} \delta s \ , \nonumber \\
& & { \delta \sigma^{-+} }
=
- { N^- \delta N^+ } \to - p_2 \delta p_1 = \frac{1}{2} \delta t \ .
\label{eq:delta_area}
\end{eqnarray}
These operations are defined only at the angle points ${x_i}$, and one has to distinguish between ``left'' and ``right'' variations, as shown in Fig. 2.

\begin{figure}[ht]
 $$\includegraphics[angle=90,width=0.5\textwidth]{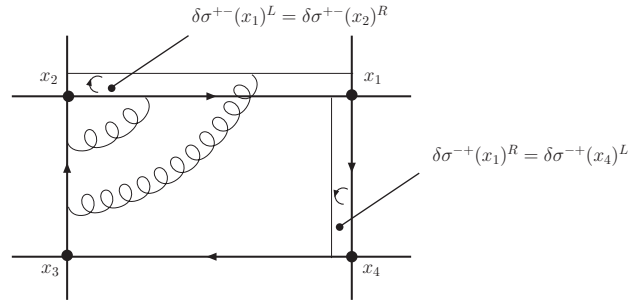}$$
   %\vspace{0.0cm}
   \caption{Infinitesimal area transformations for a light-cone rectangle in the null-plane: we consider only those area variations which conserve the angles between the sides. These variations are defined in the corners $x_i$.}
\end{figure}

${ W(\Gamma_{\Box})}$ is one of the best studied examples of the (partially) light-like objects which are known to lack multiplicative renormalizability \cite{WL_LC_rect}. Another example is provided by the transverse-momentum dependent parton density with purely light-like semi-infinite gauge links \cite{WL_LC_rapidity, CS_all}: extra divergency stems from the one-loop vertex-type graph Fig. 1(a) in covariant gauges or from the self-energy graph Fig. 1(b) in the light-cone gauge (in the large-$N_c$ limit)
\begin{eqnarray}
& &  {\rm TMD}_{\rm UV\otimes LC}
= - \frac{\alpha_s N_c}{2\pi}
 \Gamma (\epsilon) \left[ 4\pi \frac{\mu^2}{-p^2} \right]^{\epsilon} \times \nonumber \\
 & & \cdot \
 \delta (1-x) \delta^{(2)} (\bm{k}_\perp ) \ {\int_0^1\! dx \ \frac{x^{1-\epsilon}}{(1-x)^{1+\epsilon}} } \ .
\end{eqnarray}
In both cases, the reason of the renormalizability breakdown is that the light-like Wilson lines (or the seemingly {\it standard} quark self-energy in the light-cone axial gauge) are more singular than the usual Green functions. Remarkable duality between these two cases will be discussed below.

In order to decrease the power of singularity, one can follow the method proposed in \cite{CAD_universal}. With Eq. (\ref{eq:delta_area}) in mind, let us define the area logarithmic derivative on the light-cone
\begin{eqnarray}
  \frac{\delta}{\delta \ln \sigma}
  \equiv
  \sigma_{+-} \frac{\delta}{\delta \sigma_{+-}}
  +
  \sigma_{-+} \frac{\delta}{\delta \sigma_{-+}} \
  \label{eq:area_log}
\end{eqnarray}
and apply this operator to the r.h.s. of the Eq. (\ref{eq:WL_LC_1loop}):
\begin{eqnarray}
& & \frac{\delta}{\delta \ln \sigma} \ \ln W(\Gamma_\Box)
= - { \frac{\alpha_s N_c}{2\pi} } \ \frac{1}{\epsilon}  \times \\
& & \times \( \[{-2N^+N^-\mu^2  + i0}\]^\epsilon + \[{2N^+N^-\mu^2 + i0}\]^\epsilon \) \ . \nonumber
\label{eq:log_der}
\end{eqnarray}
Then the finite {cusp anomalous dimension} results from:
\begin{equation}
 \mu \frac{d}{d\mu} \frac{\delta \ \ln W(\Gamma_\Box)}{\delta \ln \sigma}
 =
 - 4 \ { \Gamma_{\rm cusp} } \ , \ {\Gamma_{\rm cusp} = \frac{\alpha_s N_c}{2 \pi} } + O(\alpha_s^2) \ .
 \label{eq:full_der}
\end{equation}
We get the finite result (\ref{eq:full_der}) by making use of the logarithmic {area derivative} (\ref{eq:area_log}), given that the infinitesimal area variations are defined as in (\ref{eq:area_derivative}). Equation (\ref{eq:full_der}) describes the dynamical properties of the light-like Wilson loops. We relate, therefore, the {geometry} of the loop space (expressed in terms of the area differentials) to the {dynamics} of the fundamental degrees of freedom---the gauge invariant, regularization independent {light-like Wilson loops}.

\section{Modified Schwinger approach and the combined evolution}

Let us show that the trick (\ref{eq:log_der}, \ref{eq:full_der}) is not just a handy technical tool, but a direct consequence of the geometrical properties of the loop space, whose constituents are the non-renormalizable cusped light-like Wilson loops. To this end, we shall start with the fundamental {quantum dynamical principle} proposed by Schwinger \cite{Schwinger51}. According to the latter, the quantum action operator $S$ governs variations of arbitrary states:
\begin{equation}
 {  \delta \langle \ \alpha' \ |\ \alpha''\  \rangle }
 =
 \frac{i}{\hbar} {  \langle \ \alpha' \ | \delta S | \ \alpha'' \ \rangle } \ .
 \label{eq:principle}
\end{equation}
However, this equation, being valid for renormalizable quantities, knows nothing about the non-renormalizable ones.

The previous results give us a clue to further analysis: let us study the area variations defined in (\ref{eq:delta_area})
\begin{equation}
 {  \frac{\delta}{\delta \sigma} \langle \ \alpha' \ |\ \alpha''\  \rangle }
 =
 \frac{i}{\hbar} {  \langle \ \alpha' \ | \frac{\delta}{\delta \sigma} S | \ \alpha'' \ \rangle } \ ,
 \label{eq:principle_mod}
\end{equation}
take into account the renormalization group invariance of the Schwinger equation in the {\it weak} form and apply the resulting operator to the cusped light-like contours.

Given that the space under consideration is made of scalar objects owning different geometrical and topological properties, one concludes that the equations of motion satisfied by those objects should prescribe the laws according to which they change their form. The motions in the loop space are, putting it formally, variations of the contours \cite{MM_WL}. Therefore, the problem arises how to find the correct form of the operator $S$, which is responsible for the form variations of the lightlike cusped loops (Wilson null-polygons).
Within the standard approach, one utilizes (\ref{eq:principle}) in the form (\ref{eq:sch_dy_YM}) and obtains the set of the MM Eqs. (\ref{eq:MM_general}). We will follow another strategy, trying to avoid using operations which implicitly assume the smoothness of the Wilson loops under consideration. For the sake of clarity, consider at first a generic Wilson loop $W (\Gamma)$ without specifying whether it is smooth or not. Its perturbative expansion reads
\begin{eqnarray}
 & & W (\Gamma) =
 W^{(0)} + W^{(1)} = \\
 & & 1 - \frac{g^2 C_F}{2} \ \oint_\Gamma \oint_\Gamma \ dz_\mu dz_\nu' \ D^{\mu\nu} (z - z') + O(g^4) \ , \nonumber
 \label{eq:W_generic_pert}
 \end{eqnarray}
where $D^{\mu\nu}$ is free dimensionally regularized ($\omega = 4 - 2 \epsilon$) gluon propagator
\begin{eqnarray}
  & & D^{\mu\nu} (z-z')
  =
  - g^{\mu\nu} \ \Delta (z - z') \ , \nonumber \\
  & & \Delta(z-z')
  =
  \frac{\Gamma(1-\epsilon)}{4\pi^2} \ \frac{(\pi \mu^2)^\epsilon}{[- (z-z')^2 + i0]^{1- \epsilon}} \ .
  \label{eq:gluon_prop_DR}
\end{eqnarray}
Here we adopt the Feynman covariant gauge and extract the scalar part of the propagator $\Delta (z)$. Important issues related to the gauge- and regularization independence of subsequent calculations will be studied elsewhere \cite{ChMVdV_2012}. Therefore, the l.h.s. of the Eq. (\ref{eq:principle_mod}), being applied to the Wilson loop (\ref{eq:W_generic_pert}), yields
\begin{eqnarray}
 & & \frac{\delta W (\Gamma)}{\delta \sigma_{\mu\nu}}  = \\
 & & = \frac{g^2 C_F}{2} \ \frac{\delta }{\delta \sigma_{\mu\nu}}  \oint_\Gamma \oint_\Gamma \ dz_\lambda dz^{'\lambda} \ \Delta (z - z') + O(g^4) \ . \nonumber
 \label{eq:W_area_diff_2}
 \end{eqnarray}
The area differentiation can be performed by making use of the Stokes theorem
\begin{eqnarray}
 & & \oint_\Gamma dz_\lambda \ {\cal O}^\lambda
 =
 \frac{1}{2} \int_\Sigma \ d\sigma_{\lambda \rho} (\partial^\lambda {\cal O}^\rho - \partial^\rho {\cal O}^\lambda) \ , \nonumber \\
 & & {\cal O}^\lambda
 =
 \oint_\Gamma dz_\lambda' \ \Delta (z) \ ,
\label{eq:W_diff_Stokes}
\end{eqnarray}
where $\Gamma$ is considered as the boundary of the surface $\Sigma$.
After simple manipulations and the path differentiation at the same point $x$, one obtains the leading order term of the Makeenko-Migdal equation (\ref{eq:MM_general}):
\begin{equation}
 \partial_\mu \frac{\delta W(\Gamma_\bigcirc)  }{\delta \sigma_{\mu\nu} (x)}
 =
 \frac{g^2N_c}{2} \oint_{\Gamma_\bigcirc} \ dy_\nu \ \delta^{(\omega)} (x -y) + O(g^4)\ .
 \label{eq:MM_LO_smooth}
\end{equation}
However, we have to be careful with this result: to derive it, one tacitly assumed that the Stokes theorem is applicable for all Wilson loops of interest. The latter is not the case in general, for that reason we denote the {\it good} (smooth enough) contours by a circle index $\Gamma_\bigcirc$. It is worth noting that in the $2D$ QCD, the area differentiation is reduced to the ordinary derivative, since the gluon propagator (\ref{eq:gluon_prop_DR}) for $\omega=2$ behaves as the logarithm of $z$, what yields\
\begin{equation}
  W(\Gamma_\bigcirc)^{\rm 2D}
  =
  \exp\[ - \frac{g^2 N_c}{2} \Sigma \] \ , \ \Sigma = \ {\rm area \ inside\ } \Gamma_\bigcirc \ ,
  \label{eq:2D_area}
\end{equation}
so that
$
  2 \ln W'_\Sigma =
  - {g^2 N_c}.
$
Calculating, in the similar manner, the next-to-leading terms, one can come to the full MM Eq. (\ref{eq:MM_general}). Nevertheless, we shall stop at this point and make a couple of steps backward, since we are interested in the loops which apparently do not satisfy the Stokes theorem conditions. For that reason, we will try to learn something about the area variations of the Wilson loops without using the Stokes theorem, but taking into account instead an explicit form of the gluon propagator (which develops a specific singularity on the light-cone), Eq. (\ref{eq:gluon_prop_DR}).
Restricting ourselves with the area variations of the type (\ref{eq:delta_area}), one obtains the area derivative of the Wilson rectangle
\begin{eqnarray}
 & & \frac{\delta W(\Gamma_\Box)  }{\delta \sigma_{\mu\nu}} = \frac{g^2C_F}{2} \frac{\Gamma(1-\epsilon) (\pi \mu^2)^\epsilon}{4\pi^2} \
 \frac{\delta }{\delta \sigma_{\mu\nu}} \sum_{i,j} (v_j^\lambda v_j^\lambda) \cdot \nonumber  \\
 & & \times \int_0^1\int_0^1\! \frac{d\tau d\tau'}{[- (x_i - x_j - \tau_i v_i + \tau_j v_j)^2 + i0]^{1 -\epsilon}} \ ,
 \label{eq:WL_rect_area}
\end{eqnarray}
where the sides of the rectangle are parameterized as $z_i^\mu = x_i^\mu - v_i^\mu \tau$ with the vectors $v_i$ having the dimension $[{\rm mass}^{-1}]$ \cite{WL_LC_rect}.
It is a remarkable feature of the light-like loops that the area dependence gets factorized from the integrals and can be evaluated explicitly (we remind that $2 (v_1v_2) = 2 N^+N^-$ in the notations of the Eq. (\ref{eq:delta_area}))
\begin{eqnarray}
& & W^{(1)}(\Gamma_\Box)
=
- \frac{\alpha_s N_c}{2\pi} {\Gamma(1-\epsilon)} (\pi \mu^2)^\epsilon \nonumber \\
& & (- 2 N^+N^-)^\epsilon\ \frac{1}{2} \int_0^1\int_0^1\! \frac{d\tau d\tau'}{[(1-\tau)\tau']^{1 -\epsilon}}   \ .
 \label{eq:WL_rect_area_NN}
\end{eqnarray}
On the other hand, light-like Wilson lines with $v_i^2 = 0$ produce additional (compared to the the off-light-cone case) singularity, which shows up as a second-order pole in $\epsilon$, while the cusps make the conformal invariance of the Wilson loop anomalous due to the presence of the {\it skewed} scalar products $(v_i v_j) \neq 0$ instead of the conformal ones $v_i^2$. Then, performing the area
$\delta / \delta \ln \sigma = \delta / \delta \ln (2 N^+N^-) $
 and the mass scale (logarithmic) differentiation of Eq. (\ref{eq:WL_rect_area_NN}) and collecting all relevant leading order terms, we come to the result
\begin{equation}
 \mu \frac{d }{d \mu } \ {\[  \frac{\delta}{\delta \ln \sigma} \ \ln \ W (\Gamma) \] }
 =
 - \sum { \Gamma_{\rm cusp} } \ ,
 \label{eq:mod_schwinger}
\end{equation}
which was anticipated in Eq. (\ref{eq:full_der}) and which is derived now as a direct consequence of the Schwinger approach. It is not surprising that this result resembles the situation in $2D$ QCD considered above. Here again the area derivative turns into the ordinary derivative for the same reason: the null-plane, where the light-like Wilson rectangles are defined, is an effectively two-dimensional space, where the set of the MM Eqs. becomes closed and--at least, in principle--solvable \cite{MM_WL, WL_Renorm}
More detailed technical discussion will be reported elsewhere \cite{ChMVdV_2012}.

Note that the r.h.s. of Eq. (\ref{eq:mod_schwinger}) is given by the cusp anomalous dimension, which is a universal (independent of the form of a contour) quantity and is known perturbatively up to the $O(\alpha_s^3)$ order.  Let us examine whether the above result is only the one-loop approximation, or it can be extended to the higher orders. To this end we take into account the linearity of the (angle-dependent) cusp anomalous dimension in the large-angle asymptotic regime with respect to the logarithm of the cusp angle $\chi \to  \frac{1}{2}\ln \frac{(2 v_i v_j)^2}{v_i^2 v_j^2}$ \cite{KR87}:
\begin{equation}
 \lim_{\chi \to \infty} \Gamma_{\rm cusp} (\chi, \alpha_s)
 =
 \sum \alpha_s^n C_n (W) a_n (W) \ \ln \frac{(2 v_i v_j)}{|v_i| |v_j^2|} \ ,
 \label{eq:cuspCD_LC}
\end{equation}
where the {\it maximally non-Abelian} numerical coefficients are
\begin{equation}
 {C_k} \sim C_F \ N_c^{k-1} \to {\frac{N_c^k}{2} } \ ,
\end{equation}
and $a_n$ are cusp-independent factors.
This regime corresponds exactly to the light-cone situation given that the angle-dependent logarithms turn into additional poles in the regularization parameter $\epsilon$: $\chi \to \frac{(v_iv_j)^\epsilon}{\epsilon}$, see Ref. \cite{WL_LC_rect, KR87}.
Namely, the area variable $ \ \sim (v_iv_j)$ enters the regularized area-dependent cusp anomalous dimension in the light-cone limit as
\begin{equation}
 \Gamma_{\rm cusp} ({\rm area}, \epsilon, \alpha_s)
 =
 \sum \alpha_s^n C_n (W) a_n (W) \ \frac{{\rm area}^\epsilon}{\epsilon} \ ,
 \label{eq:cuspCD_LC_1}
\end{equation}
and, after logarithmic area differentiation, one gets the finite perturbative expansion of the cusp anomalous dimension
\begin{equation}
 \lim_{\epsilon \to 0} \frac{d \ \Gamma_{\rm cusp} ({\rm area}, \epsilon, \alpha_s)}{d\ \ln \ {\rm area}}
 =
 \sum \alpha_s^n C_n (W) a_n (W) \ ,
 \label{eq:cuspCD_LC_2}
\end{equation}
supporting the validity of our result (\ref{eq:mod_schwinger}) in the higher orders given that, by definition, $\Gamma_{\rm cusp} =
- d \ln W / d \ln \mu$. This means that the result (\ref{eq:mod_schwinger}) should be understood as an all-order one, alike the MM Eq. (\ref{eq:MM_general}): the both are exact and non-perturbative, while the r.h.s's of each can be evaluated order by order in perturbation theory. Explicit proof of this statement will be given separately.
Let us point out that Eq. (\ref{eq:full_der}) is consistent with the non-Abelian exponentiation of the regularized  Wilson loops with cusps:
\begin{equation}
 W (\Gamma_\Box; \epsilon)
 =
 \exp \[ \sum_{k=1} \alpha_s^k \ C_k (W) F_k (W) \ \] \ ,
\end{equation}
where the summation goes over all two-particle irreducible diagrams, whose contribution is given by the {\it web} functions $F_k$ \cite{WL_expo, KR87}.
Therefore, Eq. (\ref{eq:full_der}) can be applied, in principle, for computing the higher-order perturbative corrections to the cusp anomalous dimension, given that we have a closed recursion of the  perturbative equations \cite{ChMVdV_2012}.

Note that beside the rectangular lightlike Wilson loops in the null-plane, Eq. ({\ref{eq:mod_schwinger}}) is valid for
the transverse-momentum densities with the longitudinal gauge links on the light-cone $\Phi (x, \vecc k_\perp)$, so that
\begin{equation}
  \mu \frac{d }{d \mu } \ \[ \frac{d}{d \ln \theta} \ \ln \ { \Phi (x, \vecc k_\perp) } \]
  =
  2 { \Gamma_{\rm cusp} } \ ,
\end{equation}
where the corresponding area is hidden in the rapidity cutoff $\theta \sim (N^+N^-)^{-1}$ \cite{CS_all}. Another remarkable example is given by the $\Pi$-shape loop with one (finite) segment lying on the light-cone \cite{WL_Pi}. In the one-loop order, one has in the large-$N_c$ limit
\begin{eqnarray}
  & & W(\Gamma_\Pi)
  =
  1 + \frac{\alpha_s N_c}{2\pi} \  + \nonumber \\
  & & \[ - L^2 (NN^-) + L (NN^-)  - \frac{5 \pi^2}{24}  \] \ , \\
  & & L(NN^-)
  = \frac{1}{2}\(\ln (\mu N N^- + i0) + \ln (\mu NN^- + i0) \)^2 \ , \nonumber
  \label{eq:pi_1loop}
  \end{eqnarray}
where the area is defined by the product of the light-like $N^-$ and non-light-like $N$ vectors in the coordinate space.
The $\Pi$-shaped Wilson loop (\ref{eq:pi_1loop}) obeys Eq. (\ref{eq:mod_schwinger}) as well:
\begin{equation}
  \mu \frac{d }{d \mu } \ \[ \frac{d}{d \ln \sigma} \ \ln \ { W(\Gamma_\Pi) } \]
  =
  - 2 { \Gamma_{\rm cusp} } \ ,
\end{equation}
the latter being responsible for the renormalization-group behavior of the collinear parton densities in the large-$x$ regime and for the anomalous dimensions of conformal operators with large Lorentz spin \cite{WL_Pi}.

\section{Discussion and outlook}
The quantum dynamical approach formulated by Schwinger some half a century ago provides a full and consistent description of the geometrical properties of the loop space. The Wilson loops of arbitrary shape are the fundamental degrees of freedom within this picture, and the Makeenko-Migdal equations (\ref{eq:MM_general}) can be derived from the Schwinger-Dyson set of equations for the renormalizable loops. In general, the system of the MM equations is not closed and cannot be straightforwardly applied to calculation of any useful quantity.

\begin{figure}[ht]
 $$\includegraphics[angle=90,scale=0.5]{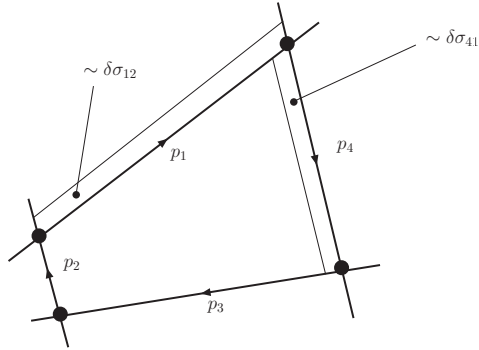}$$
   %\vspace{0.2cm}
   \caption{\label{fig:3}Generic infinitesimal area variations responsible for the conjectured quantum-dynamical loop equations for Wilson light-like $n-$polygons. Upon finding, the corresponding area differential will be reported separately.}
\end{figure}

The problem we addressed in this paper is how to construct an appropriate system of the dynamical equations valid for the cusped light-like Wilson loops, the latter having stronger singularities than their off-light-cone relatives. General solution of this problem is still lacking, but we have found that there is a class of the loops for which some simplifications are possible and helpful.
In particular, in the large-$N_c$ limit, in the case of the rectangular light-like Wilson loops defined in the null-plane $\vecc z_\perp =0$, the area functional derivative is reduced to the normal derivative for the dimensionally regularized (not renormalized) loops. The area evolution equations (which can be treated as the non-renormalizable counter-parts of the MM equations) in the coordinate space appear to be equivalent to the energy evolution equations for the cusped Wilson loops in the momentum space.
The nonperturbative nature of the dynamical loop equations enables us, in principle, to construct a chain of equations for, e.g., the cusp anomalous dimension, so that one can calculate it recursively in any given order in $\alpha_s$.

To conclude, we have taken a couple of first steps towards the understanding of the connection between the differential geometrical properties of the loop space in terms of the area evolution and the dynamics encoded in the cusps---the angles between the light-like straight lines. Within this picture, the dynamics of the elements of the loop space can be taken into account by obstructions of the (initially) smooth Wilson loops, with the obstructions playing the role of the {\it sources} within Schwinger's fields-sources picture. We have shown that the Schwinger quantum dynamical principle can be used as a guiding rule to study at least one special class of the elements of the loop space, the cusped Wilson exponentials on the light-cone. So far we have been able to implement the program only in one of the simplest cases, the rectangular contour in the transverse null-plane. In Fig. 3, a more complicated configuration is displayed: an arbitrary quadrilateral contour, the area evolution of which is less trivial and deserves a separate study. Another interesting application of the approach proposed in the present work can be found in study of the gravity Wilson loops \cite{WL_GR} and of the non-light-like Wilson polygons and the polyhedra \cite{WL_Polygons}. Evaluation of the minimal surface differentials for a variety of more complicated cusped Wilson loops defined in non-trivial manifolds is needed in order to derive corresponding area/energy evolution equations based on the quantum dynamical principle \cite{WL_min_surf}.

\noindent
{\it Acknowledgements}\\
We thank I.V. Anikin and N.G. Stefanis for careful reading of the manuscript and stimulating critical remarks. We appreciate important comments by Y.M. Makeenko.

%%%%%%%%%%%%%%%%%%%%%%%%%%%%%%%%%%%%%%%%%%%%%%%%%%%%%%%%%%%%%%%%%%%%%%%%%%%%%%%%%%%%%%%%%%%%%%%

\end{document}